\documentclass[aps,pra,twocolumn,showpacs,superscriptaddress]{revtex4-1}

\usepackage{graphicx}
\usepackage{amsmath}
\usepackage{color}
\usepackage[overlay,absolute]{textpos}
\definecolor{mygray}{gray}{0.4}
\definecolor{mylink}{rgb}{0.2, 0.2, 0.5}

\begin{document}

\begin{textblock*}{\textwidth}[0,0](19mm,11.5mm)
\footnotesize\noindent
\begin{minipage}{\textwidth}
\center
\textcolor{mygray}{Journal link: http://dx.doi.org/10.1103/PhysRevA.84.043821}
\end{minipage}
\end{textblock*}

\begin{textblock*}{0.6\textwidth}[0,0](19mm,261mm)
\footnotesize\noindent
\begin{minipage}{\textwidth}
\textcolor{mygray}{Journal ref: A. E. B. Nielsen and J. Kerckhoff, Phys.\ Rev.\ A \textbf{84}, 043821 (2011)}.\\
\textcolor{mygray}{Copyright (2011) by the American Physical Society.}
\end{minipage}
\end{textblock*}

\title{An efficient all-optical switch using a lambda atom in a cavity QED system}
\author{Anne E. B. Nielsen}
\affiliation{Max-Planck-Institut f{\"u}r Quantenoptik,
Hans-Kopfermann-Strasse 1, D-85748 Garching, Germany}
\affiliation{Lundbeck Foundation Theoretical Center for
Quantum System Research, Department of Physics and Astronomy,
Aarhus University, DK-8000 {\AA}rhus C, Denmark}
\author{Joseph Kerckhoff}
\affiliation{Edward L. Ginzton Laboratory, Stanford University, Stanford, California 94305, USA}
\altaffiliation{JILA, University of Colorado and NIST, Boulder, Colorado 80309, USA}

\begin{abstract}
We propose an all-optical switch constructed from a two-mode optical resonator containing a strongly coupled, three-state system.  The coupling allows a weak, continuous wave laser drive to incoherently control the transmission of a much stronger, continuous wave signal laser into (and through) the resonator.  We demonstrate that in this simple setup the presence of a control drive with 1/10$^{\text{th}}$ the power of the signal drive can induce near complete reflection of the signal, while its absence allows for near complete transmission.  The switch can also be operated as a set-reset relay with two control inputs that efficiently drive the switch into either the reflecting or the transmitting state.
\end{abstract}


\maketitle

\section{Introduction}\label{introduction}

While the possibility of harnessing quantum optical systems for quantum information applications has gotten much attention in recent years, the feasibility of all-optical systems for classical information processing has yet to be demonstrated, despite decades of research \cite{Mill10a}.  In order for optical logic to be competitive with silicon logic in the future, devices will have to operate on $\sim10$ attojoule energy scales (that is, hundreds of near-visible photons) \cite{Mill10b}, a regime in which quantum effects will be significant if not dominant \cite{Kerc11b}.  Practical optical devices that leverage quantum dynamics will be essential, even for completely classical information systems.

With this motivation, a device that couples a single atom (or atom-like solid-state defect) to the modes of an optical resonator has many intrinsic advantages.  Discrete atomic levels effectively `digitize' the available system states and bestow the system with an inherent nonlinearity.  Moreover, strong-coupling between atomic transitions and the optical modes may effectively route low-power lasers: a two-sided resonator that contains a single atom in an uncoupled state will transmit an on-resonant laser drive, while a coupled atom causes the laser to be reflected \cite{duankimble}.  A recently proposed all-optical set-reset relay \cite{mabuchiswitch} demonstrates that these insights may in principle be leveraged for attojoule-scale logical devices suited for solid-state, nanophotonic applications in both classical information systems and the classical information processing components of quantum information systems \cite{Kerc10}. For other recent work on all-optical switches see \cite{chen,lukin,albe,raus}.

The optical switch proposed in \cite{mabuchiswitch} is challenging to implement in practice because it requires a four level atom simultaneously coupled strongly to three cavity modes, and this raises the question whether similar behavior can be achieved in a simpler setting. In the present paper we investigate the switching performance of an analogous but less complex device consisting of an only three level lambda system coupled strongly to only two cavity modes. The two ground state levels of the lambda system correspond to the two possible states of the switch, and the excited state facilitates coupling of the lambda system to the optical fields. The presence or absence of a control field driving one cavity mode determines whether a signal field driving the other cavity mode is reflected or transmitted through the cavity. We find that a weak control field can reduce the transmitted intensity of a much stronger signal field significantly, for instance by a factor of ten for a moderately strong coupling and arbitrarily well for sufficiently strong coupling. When operated with one control field, the switch exhibits an asymmetry in the switching rates between the reflecting and the transmitting state, and we demonstrate that this asymmetry can be used to configure a dual-control, set-reset relay in the same device.

In the following, we first explain the working principle of the switch and illustrate the dynamics for a particular choice of the parameters in Sec.~\ref{basic}. We then investigate the dependence of the performance on the parameters in Sec.~\ref{parameter} and summarize the conclusions in Sec.~\ref{conclusion}.

\section{Basic functioning of the switch}\label{basic}

\begin{figure}
\includegraphics[width=\columnwidth,clip]{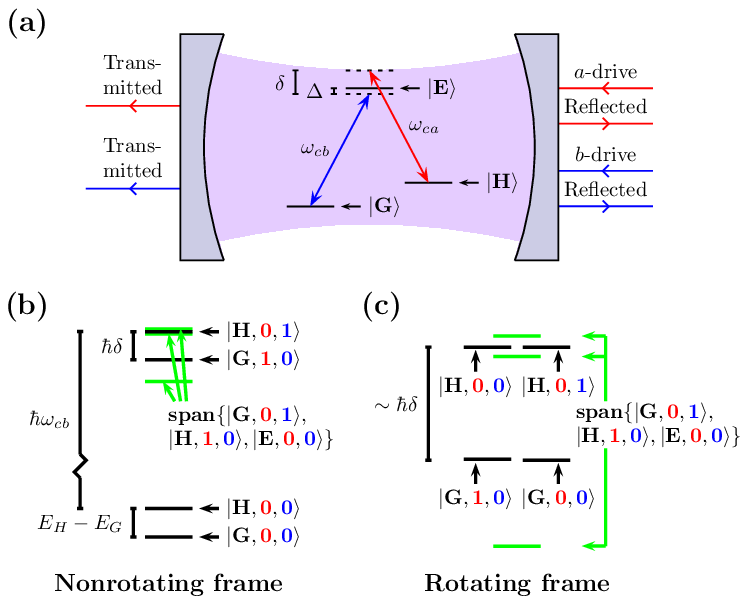}
\caption{(Color online) (a) The switch consists of a three level lambda system coupled to two cavity modes with frequencies $\omega_{ca}$ and $\omega_{cb}$, respectively. The $a$-mode ($b$-mode) is coupled to the $|H\rangle\leftrightarrow|E\rangle$ ($|G\rangle\leftrightarrow|E\rangle$) transition and spontaneous emission into radiation modes accompanied by a transition from $|E\rangle$ to $|H\rangle$ or from $|E\rangle$ to $|G\rangle$ also occurs. When one control field is used, the state (`on' or `off') of the a-drive ideally determines whether the signal field (b-drive), which is always on resonance with the bare cavity, is reflected or transmitted through the cavity, and when two control fields are used (one driving the a-mode and the other driving the b-mode at a different frequency than the signal field), the switch is ideally driven into the reflecting (transmitting) state if a few
photons are present in the control field driving the a-mode (b-mode). (b) The level structure of the seven energy eigenstates of the switch with at most one excitation in the lambda system or in one of the cavity modes when all couplings to the environment (including driving) are neglected and states with more excitations are ignored. (We use the parameters in Tab.~\ref{parameters} and draw the level structure for $\omega_{ca}=\omega_{cb}$.) The three levels for which no quantum states are given are linear combinations of $|E,0,0\rangle$, $|H,1,0\rangle$, and $|G,0,1\rangle$, where $|A,n,m\rangle\equiv|A\rangle\otimes|n\rangle_a\otimes|m\rangle_b$ and $|n/m\rangle_{a/b}$ is a photon number state of cavity mode $a$/$b$. (c) The same as (b), but in the rotating frame defined in section~\ref{theory}.} \label{setup}
\end{figure}

As illustrated in Fig.~\ref{setup}(a), the switch consists of a three level lambda system coupled to two cavity modes with annihilation operators $a$ and $b$, respectively. The lambda system could be a single atom, but it could also be a solid-state defect, for instance a nitrogen vacancy center in diamond \cite{NVC4,NVC3}, and there are likewise several possibilities for the choice of resonator. In the following, we shall keep the discussion general rather than specializing to a specific implementation.

\begin{figure}
\includegraphics[width=\columnwidth]{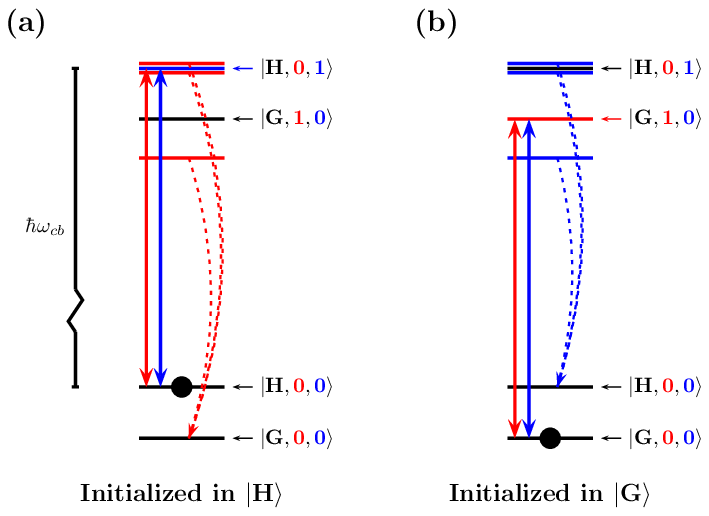}
\caption{(Color online) The response of the switch to the signal and control drives, depicted using the lowest-lying energy eigenstates from Fig.~\ref{setup}(b).  Red and blue solid arrows represent the energies of the control (red) and signal (blue) optical drives.  Horizontal red and blue lines are the energy eigenstates containing one $\sim\hbar\omega_{cb}$ quantum of excitation that can be excited near-resonantly or off-resonantly by the control and signal drives, respectively.  Dashed arrows represent decay processes that can change the state of the switch if the associated excited states are populated by the drives (see text).  (a) The response of the system when the lambda system is in $|H\rangle$: the signal resonantly excites the cavity and is transmitted; when the control drive is present, it near-resonantly excites two excited states that will efficiently pump the switch into the $|G\rangle$ state.  (b) The response of the system when the lambda system is in $|G\rangle$: the signal drive cannot efficiently excite the device and is thus reflected; however, occasional, off-resonant excitations will cause the lambda system to be pumped into the $|H\rangle$ state.  The set-reset relay configuration is not depicted.}\label{mechanism}
\end{figure}

To explain the working principle of the switch, we first assume that the $a$-mode of the cavity and the $a$-drive are both absent and that one external beam resonantly drives the $b$-mode. When the lambda system is in the $|H\rangle$ state, the model assumes that there is no interaction between the $b$-mode and the lambda system, see Fig.~\ref{setup}(a). The external signal mode hence sees an empty cavity, and when driven on resonance by a continuous wave coherent field with amplitude $\alpha_b$ (normalized such that $|\alpha_b|^2$ is the average number of photons per unit time in the beam), the steady state of the $b$-mode is a coherent state with amplitude $\sqrt{2\kappa_{b,\textrm{in}}}\alpha_b/\kappa_b$, where $2\kappa_b$ is the rate of decay of $b$-photons out of the cavity, and $2\kappa_{b,\textrm{in}}$ is the part of this decay rate coming from the input mirror, i.e., the mirror to the right in Fig.~\ref{setup}(a).

If, on the contrary, the lambda system is in $|G\rangle$, there is a strong interaction between the field in the $b$-mode and the lambda system.  In the weak driving $2\kappa_{b,\textrm{in}}|\alpha_b|^2/g_b^2\ll1$ and strong coupling $C_b\equiv g_b^2/(2\kappa_b\gamma_b)\gg1$ regime, the cavity-resonant signal drive cannot excite a large amplitude in the cavity because of the `vacuum Rabi splitting' of the energy eigenstates that arises from the strong coupling between the $b$-mode and the $|G\rangle\leftrightarrow|E\rangle$ transition. (Here, $g_b$ is the coupling strength for the interaction between the $b$-mode and the $|G\rangle\leftrightarrow|E\rangle$ transition, and $2\gamma_b$ is the decay rate from $|E\rangle$ to $|G\rangle$ due to spontaneous emission.) As a result, the steady state amplitude of the $b$-mode is reduced to $\sqrt{2\kappa_{b,\textrm{in}}}\alpha_b/(\kappa_b(1+2C_b))$ \cite{n1}.  This mechanism was proposed as a tool to implement a gate between two photonic qubits in \cite{duankimble}.

Since the amplitude of the transmitted field is $\sqrt{2\kappa_{b,\textrm{out}}}$ times the amplitude of the cavity field, where $2\kappa_{b,\textrm{out}}$ is the contribution to the cavity decay rate coming from the left cavity mirror in Fig.~\ref{setup}(a), it follows that one can control the transmission of the signal field by controlling the state of the lambda system: when the lambda system is in the $|H\rangle$ ground state, the signal is transmitted; when the lambda system is in the $|G\rangle$ ground state, very little light from the signal gets into the cavity and the signal is reflected from the cavity. We also point out that when the lambda system is in $|H\rangle$, perfect transmission is achieved for $\kappa_b=2\kappa_{b,\textrm{in}}=2\kappa_{b,\textrm{out}}$, whereas when the lambda system is in $|G\rangle$, perfect reflection occurs in the limit $C_b\rightarrow\infty$.

While the $b$-mode couples only to the $|G\rangle\leftrightarrow|E\rangle$ transition, the $a$-mode couples only to the $|H\rangle\leftrightarrow|E\rangle$ transition.  Thus, if a second field (the `control') drives only the $a$-mode (this can, e.g., be achieved through polarization selectivity) and the lambda system is in the $|H\rangle$ ground state, the control may (roughly speaking) excite an $a$-photon in the cavity if its frequency is properly tuned.  But because the $|H,1,0\rangle$ state is coherently coupled to the $|E,0,0\rangle$ state, which in turn is coherently coupled to the $|G,0,1\rangle$ state, both a lambda system excitation and a single $b$-photon are also coherently created when the $a$-drive excites the system.  The $|E,0,0\rangle$ state may spontaneously decay to $|G,0,0\rangle$, and $|G,0,1\rangle$ is transferred to $|G,0,0\rangle$ if the $b$-photon decays out of the cavity.  When either of these processes occurs, the lambda system decouples from the control drive and stays in the $|G\rangle$ state.  From the point of view of the signal field, though, the state of the switch has thus been transferred from `transmitting' ($|H\rangle$) to `reflecting' ($|G\rangle$).  The response of the device to both the signal and control drives when the lambda system is in $|H\rangle$ is depicted in Fig.~\ref{mechanism}(a) and demonstrated via simulation in section~\ref{MCsim}.

Although the signal drive cannot efficiently excite the device when the lambda system is in the $|G\rangle$ state (because of the on-resonance signal frequency and the vacuum Rabi splitting), for large but finite $C_b$ it may occasionally excite the device off-resonantly.  When this occurs, the analogous, but opposite pumping mechanisms may put the lambda system into the $|H\rangle$ state.  This re-pumping mechanism is depicted in Fig.~\ref{mechanism}(b) and also in simulation in section~\ref{MCsim}.  The trick to obtain a good switch, especially when the control field is weaker than the signal field, is to make the pumping from $|H\rangle$ to $|G\rangle$ (driven by the control drive) more efficient than the pumping in the opposite direction (driven by the signal drive) such that the lambda system spends most of its time in the $|G\rangle$ state when the control drive is on. As suggested above and we shall see in more detail below, this can be achieved if the $a$-drive near-resonantly excites the device (given its coupled energy spectrum, Figs.~\ref{setup}(b-c)), pumping the lambda system into the $|G\rangle$ state, and if the opposite pumping from $|G\rangle$ to $|H\rangle$ induced by the signal occurs only via an off-resonant excitation.

If the signal-induced pumping from $|G\rangle$ to $|H\rangle$ occurs on a time scale that is much longer than the time the switch is typically required to stay in the `reflecting' state, there is no need to keep the control drive on once it has turned the switch to `reflecting'.  To efficiently turn the switch back to `transmitting', though, one may employ a second control field driving the $b$-mode, but with a distinct optical frequency that near-resonantly excites the device, causing the lambda system to be quickly pumped to the $|H\rangle$ state. The switch hence acts as a set-reset relay, which relaxes to the `reflecting/transmitting' state by weakly driving one or the other of two control fields, while the state is fixed if both control fields are off.  This method is described in more detail in section~\ref{relaysec}.

\subsection{Theoretical model}\label{theory}

The dynamics of the system can be modeled by the master equation \cite{OSAQO}
\begin{multline}\label{meq}
\frac{d\rho}{dt}=-\frac{i}{\hbar}[H,\rho]+\mathcal{D}[\sqrt{2\gamma_a}\sigma_H]\rho
+\mathcal{D}[\sqrt{2\gamma_b}\sigma_G]\rho\\
+\mathcal{D}[\sqrt{2\kappa_a}a]\rho+\mathcal{D}[\sqrt{2\kappa_b}b]\rho,
\end{multline}
where $\rho$ is the density operator of the cavity modes and the lambda system, $t$ is time,
\begin{equation}
\mathcal{D}[c]\rho\equiv c\rho c^\dag-(c^\dag c\rho+\rho c^\dag c)/2,
\end{equation}
$\sigma_A\equiv|A\rangle\langle E|$, and parameters with a subscript $a$ are defined analogously to the corresponding parameters with a subscript $b$ (see the previous subsection), but refer to the $a$-mode and the transition between $|H\rangle$ and $|E\rangle$. The first term on the right hand side of \eqref{meq} is the Hamiltonian evolution, which includes the free evolution of the cavity modes and the lambda system, the interaction between the lambda system and the cavity modes, and the driving of the cavity, and the other terms describe spontaneous emission from $|E\rangle$ to $|H\rangle$, spontaneous emission from $|E\rangle$ to $|G\rangle$, cavity decay for the $a$-mode, and cavity decay for the $b$-mode, respectively.

In the following, we employ a standard Jaynes-Cummings model for the system's Hamiltonian and work in a rotating frame defined by the frame-boosting Hamiltonian \cite{OSAQO}
\begin{multline}
H_0=\hbar\omega_a a^\dag a+\hbar\omega_b b^\dag b +\hbar\omega_{cb}|E\rangle\langle E|\\
+\hbar(\omega_{cb}-\omega_b)|G\rangle\langle G|
+\hbar(\omega_{cb}-\omega_a)|H\rangle\langle H|,
\end{multline}
where $\omega_a$ ($\omega_b$) is the frequency of the $a$-drive ($b$-drive) and $\omega_{ca}$ ($\omega_{cb}$) is the frequency of the $a$-mode ($b$-mode) of the cavity. We likewise denote the frequency of the control field driving the $b$-mode in the set-reset relay configuration by $\omega_c$, and we use the term c-field to refer to this field. In this frame,
\begin{multline}
H=-\hbar\Theta_aa^\dag a-\hbar\Theta_bb^\dag b+\hbar\Delta |E\rangle\langle E|
+\hbar\Theta_b|G\rangle\langle G|\\
+\hbar(\Theta_a+\delta)|H\rangle\langle H|
+i\hbar g_a(a^\dag\sigma_H-a\sigma_H^\dag)\\
+i\hbar g_b(b^\dag\sigma_G-b\sigma_G^\dag)
+i\hbar\mathcal{E}_a(a^\dag-a)\\
+i\hbar\mathcal{E}_b(b^\dag-b)
+i\hbar\mathcal{E}_c(e^{-i\Omega t}b^\dag-e^{i\Omega t}b),
\end{multline}
where $\Theta_i\equiv\omega_i-\omega_{ci}$,
$\mathcal{E}_i\equiv\alpha_i\sqrt{2\kappa_{i,\textrm{in}}}$, $\alpha_c$ is the amplitude of the c-field, $\kappa_{c,\textrm{in}}\equiv\kappa_{b,\textrm{in}}$, $\Omega\equiv\omega_c-\omega_b$,
\begin{equation}
\hbar\Delta\equiv E_E-E_G-\hbar\omega_{cb},
\end{equation}
\begin{equation}
\hbar\delta\equiv E_H-E_G+\hbar(\omega_{ca}-\omega_{cb}),
\end{equation}
and $E_A$ is the energy of the state $|A\rangle$.

\subsection{Resonance condition}\label{rcon}

We first investigate the dependence of the steady state expectation value of the number of photons in the cavity modes on the frequency of the $a$-drive. In this and the following two subsections, we consider the case where the state of the switch is controlled by a single control field, and we use the parameters listed in Tab.~\ref{parameters}.  Note that the ratios of the cQED parameters in Tab.~\ref{parameters} are similar to what is achievable in modern single atom experiments \cite{Kerc11b}.  Throughout the paper, we use the quantum optics toolbox \cite{qotoolbox} for computations.

\begin{figure}
\includegraphics[width=\columnwidth,clip]{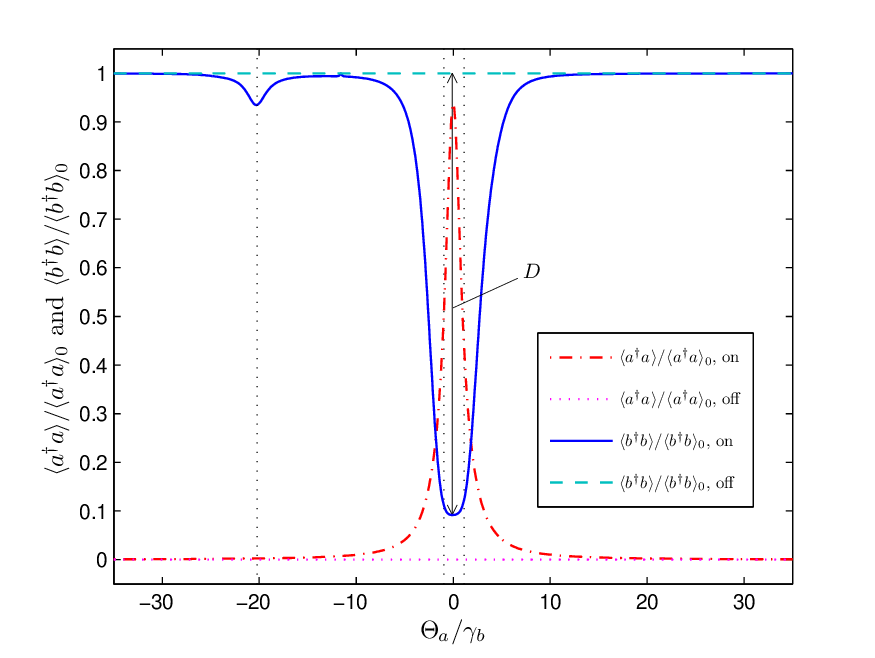}
\caption{(Color online) The expectation value of the number of photons in the cavity modes in steady state as a function of the detuning between the $a$-drive and the cavity resonance. The units $\langle a^\dag a\rangle_0$ and $\langle b^\dag b\rangle_0$ are the same expectation values for an empty cavity driven on resonance (assuming the $a$-drive is on). The labels `on' and `off' refer to the $a$-drive being on and off, respectively, and the three vertical dotted lines are the energies of the states in Fig.~\ref{setup}(c) that are linear combinations of $|E,0,0\rangle$, $|H,1,0\rangle$, and $|G,0,1\rangle$ minus the energy $\delta+\theta_a$ of $|H,0,0\rangle$ in the rotating frame. The maximum distance $D$ between the solid and the dashed lines quantifies the ability of the switch to control the transmission of the signal field.}\label{resonance}
\end{figure}

The results given in Fig.~\ref{resonance} show that a maximal decrease in the number of $b$-photons in the cavity when the $a$-drive is on is achieved if the frequency of the $a$-drive is chosen such that $\Theta_a$ is close to $-0.0915~\gamma_b$. This can be understood by considering a simplified model, in which we neglect decay processes and the driving fields and only consider states with at most one excitation in either the lambda system or one of the cavity modes. By diagonalizing the Hamiltonian, we obtain the states depicted in Fig.~\ref{setup}(c), where the three levels not labeled with an explicit state are eigenstates of the matrix
\begin{equation}\label{matrix}
\hbar\left[\begin{array}{ccc}
0 & 0 & ig_b\\
0 &\delta & ig_a\\
-ig_b & -ig_a & \Delta
\end{array}\right]
\end{equation}
written in the basis $\{|G,0,1\rangle,|H,1,0\rangle,|E,0,0\rangle\}$ (and the eigenenergies are the corresponding eigenvalues). As already described in section~\ref{basic}, if the system has been transferred to $|H,0,0\rangle$ as a result of a decay event, the $a$-drive can excite these three states (dependent on the frequency of the light) because the field drives the transition from $|H,0,0\rangle$ to $|H,1,0\rangle$. The state $|E,0,0\rangle$ can decay to $|G,0,0\rangle$ through spontaneous emission, and $|G,0,1\rangle$ can decay to $|G,0,0\rangle$ though cavity decay. Driving one of these transitions on resonance hence provides a way to drive the lambda system incoherently back to $|G\rangle$, which can be faster than the signal-driven, off resonant transfers from $|G\rangle$ to $|H\rangle$, and indeed we observe in Fig.~\ref{resonance} that the resonances in the $a$-mode coincide with the dips in the average number of $b$-photons in the cavity when the $a$-drive is on. In the following, we shall use the maximum value $D$ of $(\langle b^\dag b\rangle_\textrm{off}-\langle b^\dag b\rangle_\textrm{on})/\langle b^\dag b\rangle_0$ as a figure of merit for the switch performance, where $\langle b^\dag b\rangle_{\textrm{on}/\textrm{off}}$ is the value of $\langle b^\dag b\rangle$ in steady state when the $a$-drive is on/off. Note that this is also the relevant parameter to consider for the case of two control fields because the c-field is off except during switching from the reflecting to the transmitting state. For the parameters in Tab.~\ref{parameters}, the $a$-drive has 1/10$^{\text{th}}$ the power of the signal-drive and $D=0.908$.

\begin{table}
\begin{ruledtabular}
\begin{tabular}{cccc}
Parameter & Value & Parameter & Value\\
\hline
$\mathcal{E}_a$ & $0$ or $\sqrt{0.01}~\gamma_b$ & $g_a$ & $1.3784~\gamma_b$ \\
$\mathcal{E}_b$ & $\sqrt{0.1}~\gamma_b$ & $g_b$ & $10~\gamma_b$ \\
$\mathcal{E}_c$ & $0$ & $\Omega$ & $0$ \\
$\Theta_a$ & $-0.0915~\gamma_b$ & $\kappa_a$ & $1~\gamma_b$ \\
$\Theta_b$ & $0$ & $\kappa_b$ & $1~\gamma_b$ \\
$\delta$ & $11.5916~\gamma_b$ & $\gamma_a$ & $0.2~\gamma_b$ \\
$\Delta$ & $2.8520~\gamma_b$ & $\gamma_b$ & $1~\gamma_b$
\end{tabular}
\end{ruledtabular}
\caption{The parameters used in Figs.~\ref{setup} and \ref{mechanism} and in Secs.~\ref{rcon}-\ref{switime}. We note that $g_a^2/(2\kappa_a\gamma_a)=4.75$ and $g_b^2/(2\kappa_b\gamma_b)=50$. \label{parameters}}
\end{table}

\subsection{Monte Carlo simulations}\label{MCsim}

\begin{figure}
\includegraphics[width=\columnwidth,clip]{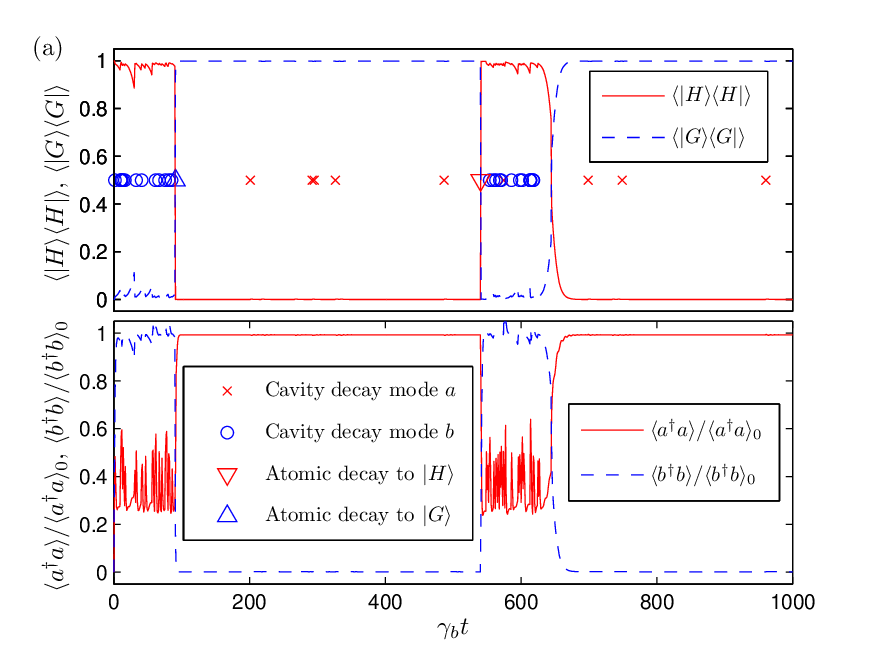}
\includegraphics[width=\columnwidth,clip]{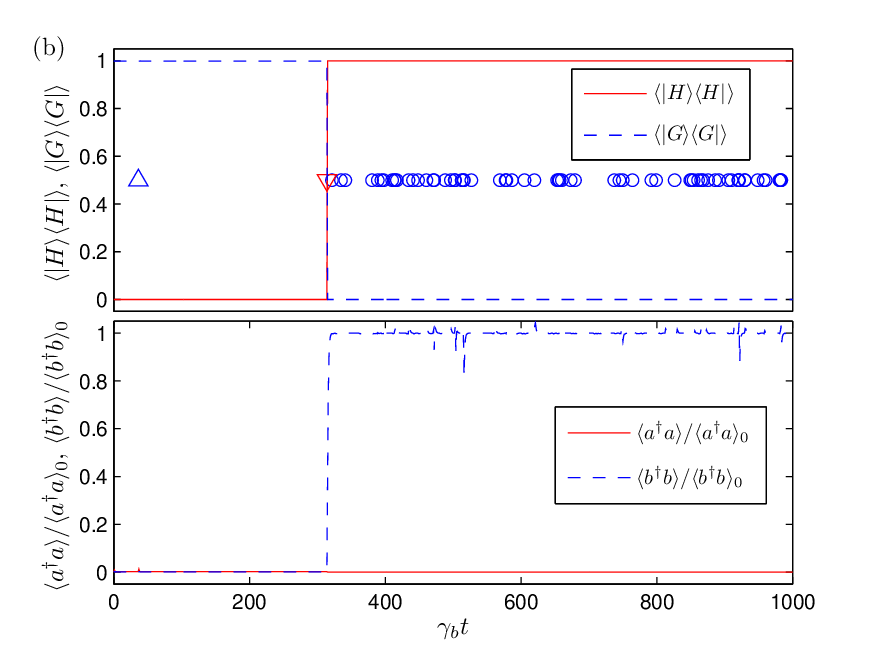}
\caption{(Color online) Particularly illustrative examples of Monte Carlo simulations of the switch dynamics for the $a$-drive being (a) on and (b) off, respectively. The initial state is $|H,0,0\rangle$ in (a) and $|G,0,0\rangle$ in (b). The crosses, circles, and triangles show the time at which the various types of jumps occur as specified in the legend in part (a) of the figure (for the cavity decays, we show only the jumps corresponding to photons transmitted through the cavity).} \label{MonteCarlo}
\end{figure}

More insight into the dynamics of the switch can be achieved through Monte Carlo simulations \cite{montecarlo}. Examples of trajectories are shown in Fig.~\ref{MonteCarlo}, where the jump operators are chosen as $\sqrt{2\gamma_a}\sigma_H$, $\sqrt{2\gamma_b}\sigma_G$, $\sqrt{2\kappa_{a,\textrm{in}}}a$, $\sqrt{2\kappa_{b,\textrm{in}}}b$, $\sqrt{2\kappa_{a,\textrm{out}}}a$, and $\sqrt{2\kappa_{b,\textrm{out}}}b$ and we assume a symmetric and lossless cavity, i.e., $\kappa_a=2\kappa_{a,\textrm{in}}=2\kappa_{a,\textrm{out}}$ and $\kappa_b=2\kappa_{b,\textrm{in}}=2\kappa_{b,\textrm{out}}$. In the case, where the $a$-drive is on, it is seen that there are two metastable states. In one of them, $\langle|H\rangle\langle H|\rangle$ and $\langle b^\dag b\rangle/\langle b^\dag b\rangle_0$ are both close to unity, while $\langle a^\dag a\rangle/\langle a^\dag a\rangle_0$ is reduced somewhat below unity. The high value of the number of $b$-photons in the cavity leads to frequent cavity decay events for the $b$-mode. In the other state, $\langle|G\rangle\langle G|\rangle$ and $\langle a^\dag a\rangle/\langle a^\dag a\rangle_0$ are close to unity, while $\langle b^\dag b\rangle/\langle b^\dag b\rangle_0$ is very close to zero. Cavity decay events occur for the $a$-mode, but at a smaller rate than for the $b$-mode in the first state because the control field is weaker than the signal field.

The specific trajectory shown in the figure provides an example of a transition from the state with large $\langle|H\rangle\langle H|\rangle$ to the state with large $\langle|G\rangle\langle G|\rangle$ via spontaneous emission and also an example where the transition occurs via absorption of an $a$-photon followed by emission of a $b$-photon into the cavity mode and out one of the mirrors. The trajectory furthermore illustrates the fact that transfers from the state with large $\langle|G\rangle\langle G|\rangle$ to the state with large $\langle|H\rangle\langle H|\rangle$ occasionally take place. On average, the switch does, however, spend more time in the desired state with large $\langle|G\rangle\langle G|\rangle$ than in the undesired state with large $\langle|H\rangle\langle H|\rangle$, and the signal field is hence mostly reflected from the cavity.

In the case, where the $a$-drive is off, there are again two states with large $\langle|H\rangle\langle H|\rangle$ and large $\langle|G\rangle\langle G|\rangle$, respectively. If the lambda system starts in $|G\rangle$, it will stay there for a while until a decay event occurs that brings the lambda system to $|H\rangle$. After the transition, the system will stay in the state with $\langle|H\rangle\langle H|\rangle=1$ as long as the control field is off because transfer from $|H\rangle$ to $|G\rangle$ is impossible, when there are no $a$-photons present.

\subsection{Switching times}\label{switime}

\begin{figure}
\includegraphics[width=\columnwidth]{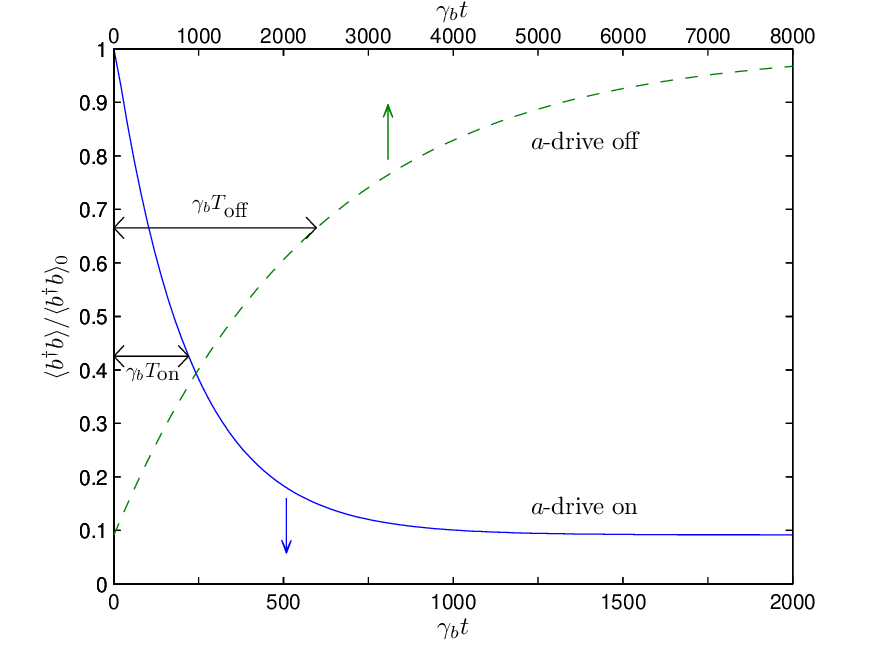}
\caption{(Color online) Time evolution of $\langle b^\dag b\rangle/\langle b^\dag b\rangle_0$ when the $a$-drive is suddenly turned on after being off for a long time (solid line) and when the $a$-drive is suddenly turned off after being on for a long time (dashed line). (Note the different scales.)}\label{switchrate}
\end{figure}

Apart from a value of $D$ close to unity, it is also important that the state of the switch changes quickly when the state of the $a$-drive changes, i.e., the presence or absence of only a few photons in the control beam should ideally suffice to cause the switch to change state. In Fig.~\ref{switchrate}, we plot the time evolution of $\langle b^\dag b\rangle/\langle b^\dag b\rangle_0$ after suddenly turning the $a$-drive on or off at time $t=0$. For the case, where the $a$-drive is turned on, we define the switching time $T_{\textrm{on}}$ as the time at which $\langle b^\dag b\rangle=(\langle b^\dag b\rangle_\textrm{off}-\langle b^\dag b\rangle_\textrm{on})e^{-1}+\langle b^\dag b\rangle_\textrm{on}$, and for the case, where the $a$-drive is turned off, we define the switching time $T_{\textrm{off}}$ as the time at which $\langle b^\dag b\rangle=(\langle b^\dag b\rangle_\textrm{on}-\langle b^\dag b\rangle_\textrm{off})e^{-1}+\langle b^\dag b\rangle_\textrm{off}$. For the considered parameters, $\gamma_bT_\textrm{on}=220$ and $\gamma_bT_\textrm{off}=2.39\times10^3$. Assuming $\kappa_a=2\kappa_{a,\textrm{in}}$, $a$-photons arrive at the cavity at the rate $|\alpha_a|^2=0.01~\gamma_b$ if the $a$-drive is on, and the switch hence changes from the transmitting state to the reflecting state after incidence of only a few $a$-photons and a few tens of $b$-photons. The opposite transition is slower and requires the absence of a few tens of $a$-photons, and the presence of a few hundred $b$-photons for these parameters. The energy required to change the state of the switch is thus in the $<100$~aJ range for both cases for near infrared photons.

\subsection{Set-reset relay}\label{relaysec}

\begin{figure}
\includegraphics[width=\columnwidth]{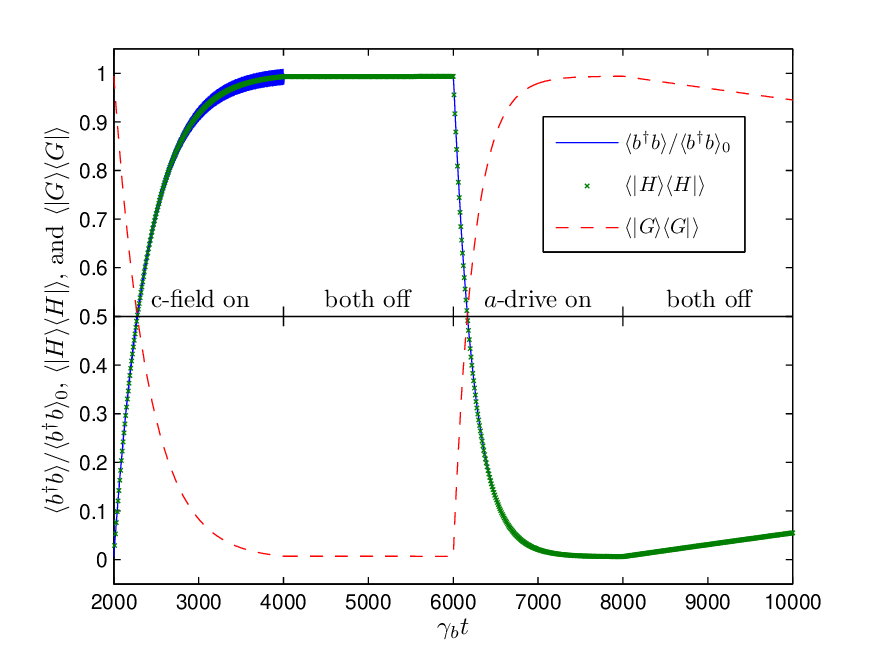}
\caption{(Color online) Dynamics of the set-reset relay. The system is initiated to $|H,0,0\rangle$ at time $\gamma_bt=0$, and the $a$-drive is turned on from $\gamma_bt=0$ to $\gamma_bt=2000$ to drive the relay to the reflecting state (not shown). From $\gamma_bt=2000$ to $\gamma_bt=4000$, the c-field is on and the $a$-drive is off, and this affects a transition to the transmitting state. The relay stays in this state when both control fields are subsequently turned off. Turning on the $a$-drive from $\gamma_bt=6000$ to $\gamma_bt=8000$ leads to a transition to the reflecting state. When the system is in the reflecting state and both control fields are off, there is a decay towards the transmitting state at the rate $1/T_\textrm{off}$, which is much slower than the transition rate seen for $\gamma_bt\in[2000,4000]$. Note that a time interval of $2000~\gamma_b^{-1}$ corresponds to the arrival of 20 control photons ($a$ or c) on average if $\kappa_{a,\textrm{in}}=\kappa_a/2=\gamma_b/2$ and $\kappa_{b,\textrm{in}}=\kappa_b/2=\gamma_b/2$.}\label{relay}
\end{figure}

As we shall see in the next section, higher values of $D$ and $T_\textrm{off}$ are achieved by increasing $g_b$. For sufficiently large $T_\textrm{off}$, we do not need continuous optical pumping to keep the switch in the reflecting state, and the switch can be operated as a set-reset relay: the $a$-beam drives the system into the reflecting state as before, but an additional $c$-beam, driving the $b$-cavity mode, can resonantly drive the system into the transmitting state. The dynamics in this case is exemplified in Fig.~\ref{relay}, where we show the results of integrating \eqref{meq} for the parameters $\Theta_a=-0.0565~\gamma_b$, $\Theta_b=0$, $\Delta=0.208~\gamma_b$, $\delta=40.1~\gamma_b$, $\mathcal{E}_a=0$ or $\sqrt{0.01\kappa_a\gamma_b}$, $\mathcal{E}_b=\sqrt{0.1\kappa_b\gamma_b}$, $\mathcal{E}_c=0$ or $\sqrt{0.01\kappa_b\gamma_b}$, $g_a=1.57~\gamma_b$, $g_b=40~\gamma_b$, $\Omega=40~\gamma_b$, $\kappa_a=\kappa_b=\gamma_b$, and $\gamma_a=0.2~\gamma_b$. As explained in the caption, the figure shows a fast transfer to the transmitting state when the c-field is turned on, a fast transfer to the reflecting state when the $a$-drive is turned on, and a slow decay towards the transmitting state when both control fields are turned off.   The decay occurs on the time scale $T_\textrm{off}$ and can be further suppressed by choosing a higher value of $g_b$.  Thus, resonantly driving the system with the $c$-beam can greatly decrease the time and energy necessary to drive the system back into the transmitting state, so that systems with large $D$ and $T_\textrm{off}$ may still be switched quickly and efficiently.

\section{Parameter dependence}\label{parameter}

The model contains several parameters, and to make an investigation of the parameter dependence feasible, we fix some of them and optimize others. We shall always choose the signal field to be on resonance with the cavity mode (i.e., $\Theta_b=0$) to ensure maximal transmission of the signal when the switch is in the transmitting state. We would like to optimize the switch performance for given amplitudes of the signal and control fields, and we hence let $\mathcal{E}_a$ ($\mathcal{E}_b$,$\mathcal{E}_c$) scale with $\sqrt{\kappa_{a,\textrm{in}}}$ ($\sqrt{\kappa_{b,\textrm{in}}}$,$\sqrt{\kappa_{b,\textrm{in}}}$) and assume $\kappa_{a,\textrm{in}}$ ($\kappa_{b,\textrm{in}}$) to scale with $\kappa_a$ ($\kappa_b$). Specifically, we choose $\mathcal{E}_a=0$ or $\sqrt{0.01\kappa_a\gamma_b}$, $\mathcal{E}_b=\sqrt{0.1\kappa_b\gamma_b}$, $\mathcal{E}_c=0$ or $\sqrt{0.01\kappa_b\gamma_b}$, and $\kappa_a=\kappa_b$ such that the intensity of the signal field is ten times larger than the intensity of the control fields for $2\kappa_{a,\textrm{in}}/\kappa_a=2\kappa_{b,\textrm{in}}/\kappa_b$. Since it is important to hit the resonances of the system, we shall optimize $\Theta_a$, $\delta$, and $\Delta$ numerically for each choice of parameters. We shall also optimize $g_a$ numerically because it turns out that the optimum is at an intermediate value of $g_a$ rather than at zero or infinity. Using $\gamma_b$ as the unit, we are then left with the parameters $\gamma_a$, $\kappa_a=\kappa_b$, and $g_b$ for the case where $\mathcal{E}_c=0$.

\begin{figure}
\includegraphics[width=\columnwidth]{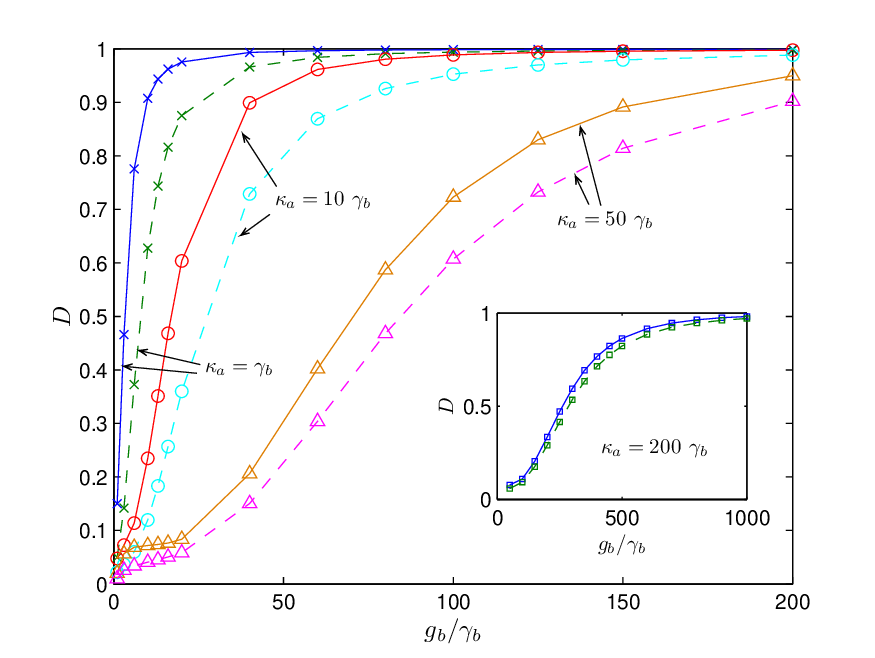}
\caption{(Color online) Signaling contrast $D$ (see Fig.~\ref{resonance}) as a function of $g_b$ for $\gamma_a=0.2~\gamma_b$ (solid lines) and $\gamma_a=\gamma_b$ (dashed lines) and various values of $\kappa_a=\kappa_b$ as indicated. $\mathcal{E}_a=\sqrt{0.01\kappa_a\gamma_b}$, $\mathcal{E}_b=\sqrt{0.1\kappa_b\gamma_b}$, $\Theta_b=0$, and $\Theta_a$, $g_a$, $\delta$, and $\Delta$ are optimized numerically to maximize $D$.}\label{gbD}
\end{figure}

In Fig.~\ref{gbD}, we provide examples of optimized values of $D$ for various choices of the parameters. For fixed $g_b$, we observe that $D$ decreases when $\kappa_a=\kappa_b$ or $\gamma_a$ is increased, whereas $D$ increases with $g_b$. This suggests that strong coupling generally leads to higher values of $D$, but the coupling for the $a$-mode should be adjusted appropriately depending on the coupling for the $b$-mode. We also observe that $D$ can be very close to unity.

\begin{figure}
\includegraphics[width=\columnwidth]{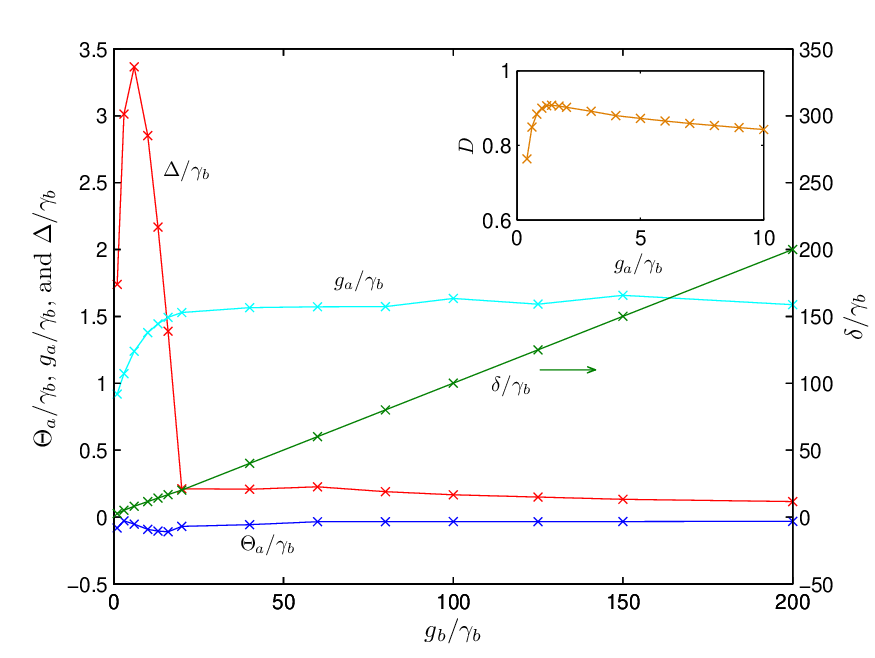}
\caption{(Color online) The optimal choice of $\Theta_a$, $g_a$, $\delta$, and $\Delta$ for the curve in Fig.~\ref{gbD} with $\gamma_a=0.2~\gamma_b$ and $\kappa_a=\kappa_b=\gamma_b$. The inset shows $D$ as a function of $g_a$ for $\gamma_a=0.2~\gamma_b$, $\kappa_a=\kappa_b=\gamma_b$, and $g_b=10~\gamma_b$ when only $\Delta$, $\delta$, and $\theta_a$ are optimized.}\label{optpar}
\end{figure}

Figure \ref{optpar} shows an example of the optimal values of the optimized parameters. Except for the very left part of the figure, we have $\delta\approx g_b$, while $g_a$, $\Theta_a$, and $\Delta$ are significantly smaller. We note that the eigenvalues of \eqref{matrix} are approximately $\pm \hbar g_b$ and $\hbar\delta$ in this regime, and hence two of the resonances in the simplified model approximately coincide. The inset shows that the precise value of $g_a$ is not critical for the performance, which is a significant experimental advantage. In general, we find that intermediate values of the parameters are typically preferred compared to hard limits, and this makes it more difficult to derive approximate analytical expressions for the quantities of interest. Finally, we note that it is important to avoid $\Theta_a+\delta=0$. If $\Theta_a+\delta=0$, the Raman transition between $|G\rangle$ and $|H\rangle$ is on resonance, and the lambda system quickly evolves into a dark state, which does not interact with the cavity modes regardless of the amplitude driving them. Specifically, the steady state of the system takes the form
\begin{equation}
|\psi_{\textrm{DS}}\rangle=(c_H|H\rangle+c_G|G\rangle) \otimes|\xi_a\rangle\otimes|\xi_b\rangle,
\end{equation}
where $|\xi_a\rangle$ ($|\xi_b\rangle$) is a coherent state with amplitude $\xi_a=\mathcal{E}_a/(\kappa_a-i\Theta_a)$ ($\xi_b=\mathcal{E}_b/\kappa_b$) and $c_H/c_G=-g_b\xi_b/(g_a\xi_a)$.

\begin{figure}
\includegraphics[width=\columnwidth]{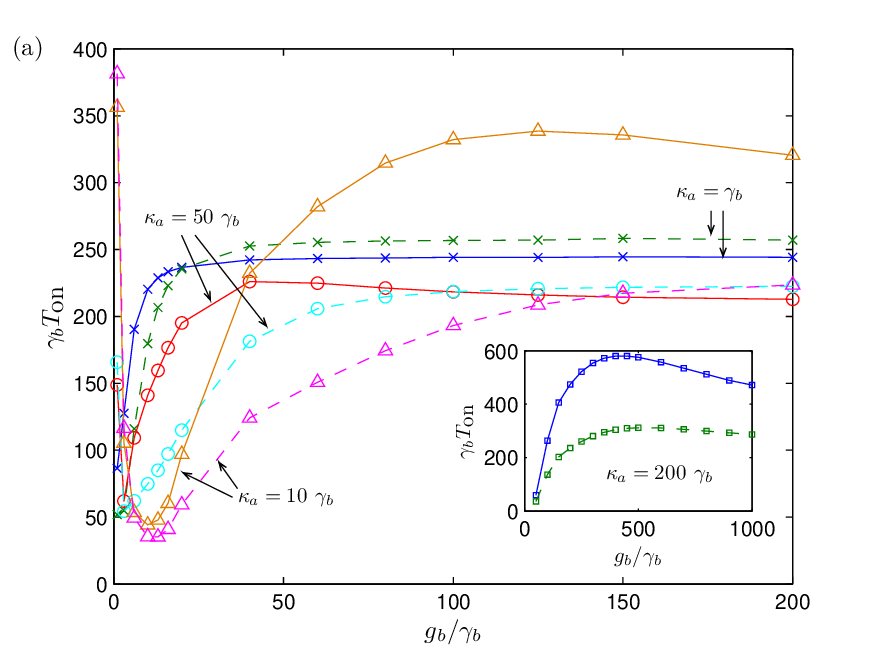}
\includegraphics[width=\columnwidth]{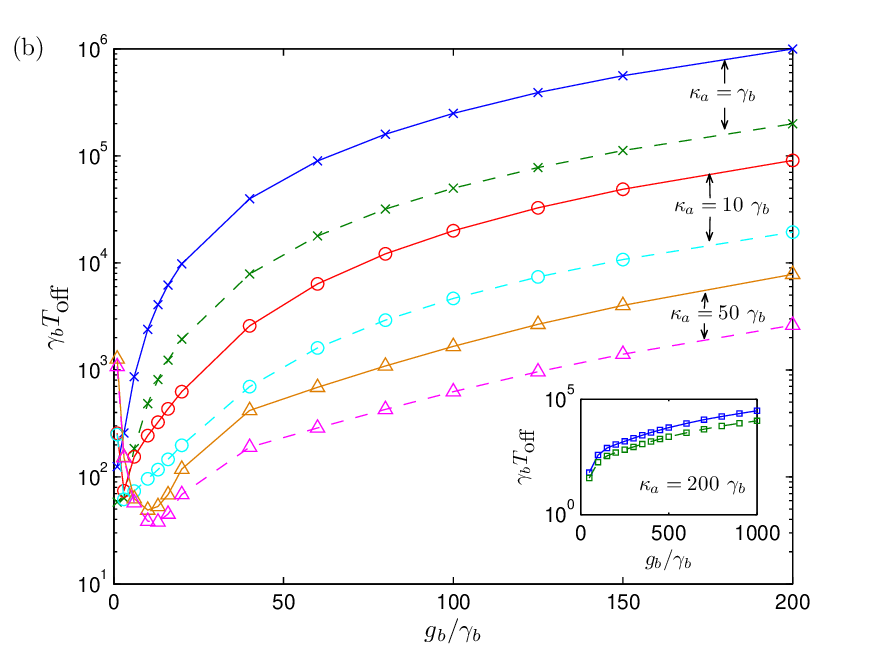}
\caption{(Color online) Switching times $T_\textrm{on}$ (a) and $T_\textrm{off}$ (b) for the same parameters as in Fig.~\ref{gbD}. Again, $\gamma_a=0.2~\gamma_b$ for the solid curves and $\gamma_a=\gamma_b$ for the dashed curves. (Note that the vertical axis is logarithmic in (b) but not in (a).)}\label{Tfig}
\end{figure}

\begin{figure}
\includegraphics[width=\columnwidth]{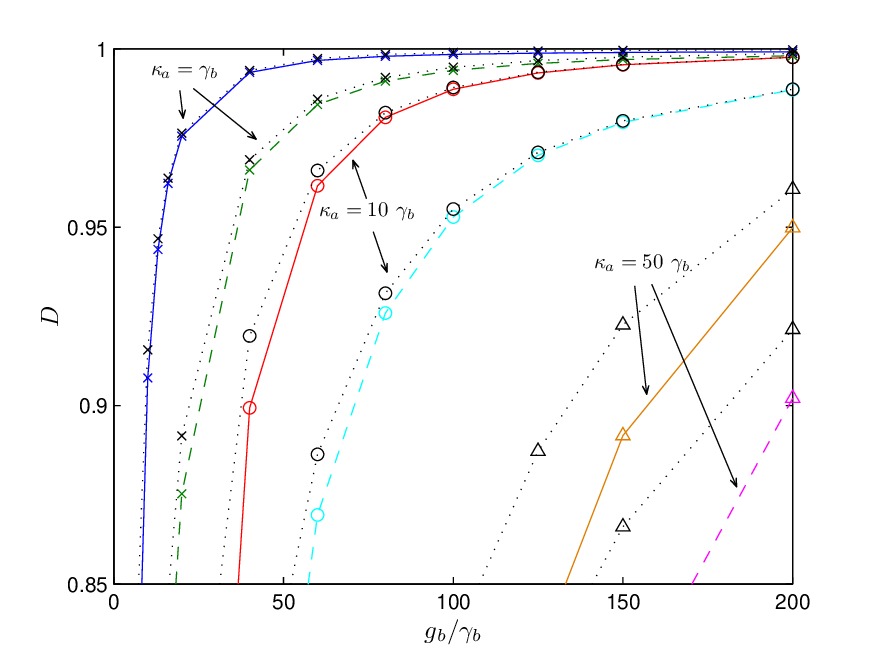}
\caption{(Color online) Enlarged view of Fig.~\ref{gbD}. The dotted lines show $T_\textrm{off}/(T_\textrm{on}+T_\textrm{off})$.}\label{Dapp}
\end{figure}

The switching times without the c-field present are shown in Fig.~\ref{Tfig}, and it is seen that switching from the transmitting state to the reflecting state generally requires only a few incident control photons. The switching times in the opposite direction tend to increase when $D$ approaches unity. In fact, there is a close connection between the switching times and $D$ for the considered parameters when $D$ is close to unity. This can be understood from the dynamics in Fig.~\ref{MonteCarlo}. Since $\langle b^\dag b\rangle_\textrm{off}$ is equal to $\langle b^\dag b\rangle_0$ for $\Theta_b=0$, $D$ is determined by the value of $\langle b^\dag b\rangle_\textrm{on}/\langle b^\dag b\rangle_0$, and since the switch jumps between two states with $\langle b^\dag b\rangle/\langle b^\dag b\rangle_0\approx1$ and $\langle b^\dag b\rangle/\langle b^\dag b\rangle_0\approx0$ when the control field is on, the value of $\langle b^\dag b\rangle_\textrm{on}/\langle b^\dag b\rangle_0$ is approximately the relative amount of time the system spends in the state with $\langle\vert H\rangle\langle H\vert\rangle\approx\langle b^\dag b\rangle/\langle b^\dag b\rangle_0\approx1$. The average time the switch stays in the transmitting state before jumping to the reflecting state is $T_\textrm{on}$, and a first estimate for the average time the system stays in the reflecting state before jumping back to the transmitting state is $T_\textrm{off}$ (this amounts to the approximation that the presence of the control field does not significantly alter the transition time for jumps from the reflecting to the transmitting state). We thus expect $\langle b^\dag b\rangle_\textrm{on}/\langle b^\dag b\rangle_0\approx T_\textrm{on}/(T_\textrm{on}+T_\textrm{off})$ and hence $D\approx T_\textrm{off}/(T_\textrm{on}+T_\textrm{off})$. This relation is checked in Fig.~\ref{Dapp}, and we find good agreement when $D$ is close to unity.

The validity of the relation suggests that high $D$ is always accompanied by high $T_\textrm{off}$. In the set-reset configuration, an efficient pathway from $|G\rangle$ to $|H\rangle$ is created by using a second control field to resonantly drive a transition between $|G,0,0\rangle$ and an excited state that can decay to $|H,0,0\rangle$ (Fig.~\ref{relay}), and it is only desirable that $T_\textrm{off}$ is very large to eliminate the need to continuously pump the lambda system to $|G\rangle$ when the switch is in the reflecting state. In this way, high $D$ and low switching times in both directions are possible in the set-reset configuration if $g_b$ is sufficiently large.

\section{Conclusion}\label{conclusion}

In conclusion, we have proposed and analyzed an attojoule-scale, all-optical digital switch comprised of a two-mode resonator coupling to a three level lambda system.  Utilizing the rich structure of the system's lowest levels of excitation (Fig.~\ref{setup}(b)), the device may be operated in an on/off-mode (using a single control field) or as a set-reset relay (using two control fields), and requires a significantly simpler system than a similar, recently proposed set-reset device \cite{mabuchiswitch}.  We have highlighted a set of configurations in which the control field(s) may route a signal beam that is 10 times more powerful, even when the device couples to both drive modes equally ($\kappa_{a,\textrm{in}} = \kappa_{b,\textrm{in}}$), and have found that efficient transmission or reflection is achievable as the lambda system's coupling to the signal mode increases.  Moreover, the device requirements for excellent controllability do not seem excessive.  For instance, projected cQED parameters $(g,\kappa,\gamma)/2\pi = (2.25,0.16,0.013)$~GHz using GaP nanophotonic resonators and diamond nitrogen-vacancy centers \cite{NVC4} (which have a natural lambda multilevel structure) are in principle sufficient for essentially perfect reflection and transmission control (the $(g_b,\kappa) = (175,10)~\gamma_b$ point in Fig.~\ref{gbD}).

Various parameters may be adjusted to optimize aspects of the dynamics such as the visibility of the switching states ($D$), the switching times and on/off switching time asymmetry.  In all configurations considered here, only a handful of photons in the control field(s) are scattered during switching events.  In the near infrared regime, this corresponds to $\lesssim$10~aJ-scale optical logic.  We emphasize that although these devices are manifestly quantum mechanical systems that rely on both the discreteness of the internal Hilbert spaces and the coherent mixing of low-lying energy eigenstates (in the `natural' basis of an un-coupled system), and involve countable numbers of energy quanta, the switching processes are principally dissipative, the field states remain essentially coherent at all times, and the devices are robustly-suited for classical logic applications.  As complex systems engineering looks towards an emerging ultra-low energy standard \cite{Mill10b}, `quasi-classical' devices like these are likely to be a critical aspect of future optical engineering.

\begin{acknowledgments}
This work has been supported by DARPA under Award No. N66001-11-1-4106, The Carlsberg Foundation, and The Danish Ministry of Science, Technology, and Innovation.
\end{acknowledgments}

\end{document}